\documentclass[twocolumn,english,aps,amsmath,prd,floatfix]{revtex4}
\usepackage[T1]{fontenc}
\usepackage[latin9]{inputenc}
\usepackage{graphicx}

\newcommand{\lyxdot}{.}

\usepackage{babel}

\begin{document}

\title{Methods for detection and characterization of signals in noisy data
with the Hilbert-Huang Transform}
\author{Alexander Stroeer}
\email{Alexander.Stroeer@nasa.gov}
\altaffiliation[Also at ]{CRESST, Department of Astronomy, University of Maryland, College Park, Maryland 20742}
\author{John K. Cannizzo}
\altaffiliation[Also at ]{CRESST, Physics Department, University of Maryland, Baltimore County, Baltimore, Maryland 21250}
\author{Jordan B. Camp}
\affiliation{Laboratory for Gravitational Physics, Goddard Space Flight Center, Greenbelt, Maryland 20771}
\author{Nicolas Gagarin}
\affiliation{Starodub, Inc., 3504 Littledale Road, Kensington, MD, 20895}
\date{\today}

\begin{abstract}
The Hilbert-Huang Transform is a novel, adaptive approach to time
series analysis that does not make assumptions about the data form.
Its adaptive, local character allows the decomposition of non-stationary
signals with hightime-frequency resolution but also renders it susceptible
to degradation from noise. We show that complementing the HHT with
techniques such as zero-phase filtering, kernel density estimation
and Fourier analysis allows it to be used effectively to detect and
characterize signals with low signal to noise ratio. 
\end{abstract}
\maketitle

\section{Introduction}

The Hilbert-Huang Transform (HHT) \cite{huang1998emd,huang2005hht}
is a novel data analysis algorithm that adaptively decomposes time
series data and derives the instantaneous amplitude (IA) and instantaneous
frequency (IF) of oscillating signals. Because this transform operates
locally on the data, and not as an integral in time over pre-selected
basis functions, it can effectively decompose non-linear, non-stationary
signals, and it is not limited by time-frequency uncertainty. Applications
of the HHT include monitoring of heart rates\cite{echeverria2001aem},
integrity of structures \cite{loutridis2004ddg}, and searching for
gravitational waves \cite{camp2007ahh}.

The HHT proceeds in two steps \cite{huang2005hht}. The first part
of the algorithm, the empirical mode decomposition (EMD), decomposes
the data into intrinsic mode functions (IMF), each representing a
locally monochromatic frequency scale of the data, with the original
data recovered by summing over all IMFs. EMD involves forming an envelope
about the data maxima and minima with the use of a cubic spline, then
taking the average of the two envelopes, and subtracting that from
the time series to obtain the residual. An iteration of this procedure
converges to an IMF, after which it is subtracted from the time series,
and the procedure begins again. The second part applies the Hilbert
transform to each individual IMF to construct an analytical complex
time series representation. The instantaneous frequency of the original
IMF is obtained by taking the derivative of the argument of the complex
time series, and the instantaneous amplitude by taking the magnitude.

Many applications of the HHT to date have involved the decomposition
of complicated mixings of non-stationary features, which may also
be frequency modulated, but these generally have not been limited
by low signal strength relative to the noise background. A different
class of problems involves signal detection and characterization at
low signal to noise ratio (SNR). The SNR of a signal $h$, as recorded
discretely according to a sampling frequency with the individual time
instances denoted by the subscript $i$, in white noise with standard
deviation $\sigma_{n}$ is defined as (matched filter definition):

\begin{equation}
SNR=\sqrt{\sum_{i}h_{i}^{2}}/\sigma_{n}\label{eq:SNR}\end{equation}

An interesting question is the effectiveness of the HHT decomposition
for low SNR. This is influenced by what we describe as intrinsic and
extrinsic effects. Intrinsic uncertainties are evident in the presence
of noise within the bandwidth of the actual signal, so that the true
waveform of the signal is never visible to the data analysis method.
Extrinsic uncertainties are induced by the data analysis algorithm
in the form of errors in the processing of the data stream due to
noise either inside or outside the signal bandwidth, leading to envelope
undershoot or overshoot, with the error possibly magnified by the
EMD iterations. Additional extrinsic uncertainties can be introduced
in the application of the Hilbert transform if the IMF is not perfectly
locally monochromatic, or due to limitations described in Bedrosian
and Nuttal theorems \cite{HuangIFAADA}; or in the determination of
the IF, as the numerical derivative of the instantaneous phase may
be subject to uncertainties and error propagation.

The length of the signal is also an important consideration in the
accuracy of the HHT decomposition. The local character of the HHT
implies a direct sensitivity of the decomposition to the local signal
amplitude relative to the noise ($I\! A/\sigma_{n}$). For a given
SNR, the signal amplitude relative to the noise increases as the signal
becomes shorter in time (see Eq. \ref{eq:SNR}). Thus shorter signals
at a given SNR will be less subject to uncertainties, and more easily
detected.

We consider in this paper methods for enhancing the HHT performance
in detecting and characterizing signals at low SNR ($<$20), and with
duration $<$100 msec. Our principal motivation for this is in analyzing
data from the Laser Interferometer Gravitational-Wave Observatory
(LIGO)\cite{barish1999lad,sigg2006sld}. Gravitational wave signals
at the current sensitivity of LIGO are expected to show only low SNR
with predicted event rates not exceeding a few per year \cite{barish1999lad}.
The adaptive and high time-frequency resolution features of the HHT
are well-suited to LIGO analysis\cite{camp2007ahh}, but its low SNR
performance remains a key issue of investigation. We focus in this
paper on simulations with time series data composed of stationary
white Gaussian noise and low SNR signals well separated in time. We
note that results are easily transferable to the general field of
low SNR analyses.

Below we present methods to limit extrinsic and intrinsic uncertainties.
We introduce a two-stage use of the HHT for detection and characterization:
detection strategies scan for excess signal power above the noise
floor, and characterization strategies extract information about the
signal frequency and power evolution in time using information from
the detection stage. To enhance the effectiveness of the HHT at low
SNR, we present the application of a number of techniques including
least squares velocity filters\cite{frei1999lsa}, Bayesian blocks
\cite{mcnabb2004obe}, zero-phase high order Finite Impulse Response
(FIR) filtering techniques \cite{smith:sae}, the fast Fourier transform
(FFT)\cite{smith:sae} and weighted adaptive kernel density estimates
\cite{silverman1986des}. We use the Ensemble EMD method \cite{WuEEMD2008}
as a tool to minimize extrinsic noise from envelope over/undershoots,
and also as a guide to an approach to estimate the uncertainty of
the decomposed frequency evolution of the signal.

\begin{figure}
{\resizebox{\hsize}{!}{\includegraphics{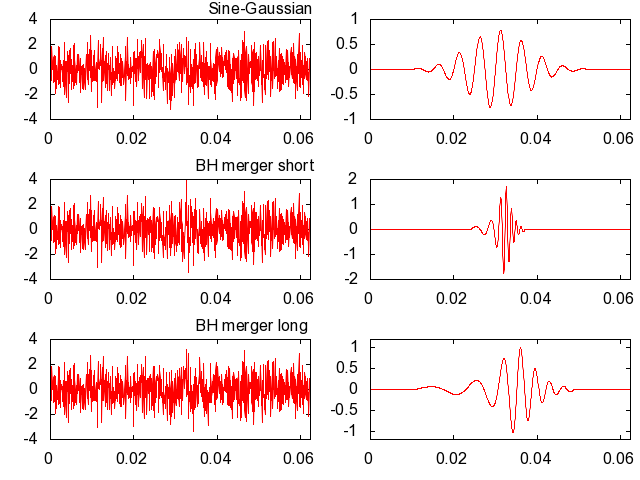}}}

\caption[Data]{Three test signals at SNR=8 are used in this paper to illustrate
our approaches - a Sine-Gaussian at 200 Hz with Q of 9 in the top
panels and a numerical simulation of a black hole binary merger (referred
to as BH merger, for details see text) in the middle (total mass low,
short duration) and bottom (total mass high, long duration) panels.
We show the signals (right side), and the signals injected in white
Gaussian noise at SNR=8 (left side).}

\label{f0} 
\end{figure}

As the HHT is an empirical technique that is not yet supported by
a fully independent theoretical basis, we test our approaches through
numerical simulations. Throughout this paper we use three test signals
to demonstrate our proposed methods (Fig.~\ref{f0}): a Sine-Gaussian
at $f_{0}=200$ Hz with Q of 9 ($h_{S\! G}=A\sin(2\pi f_{0}t_{i})\exp(-(2\pi f_{0}(t_{i}-t_{o}))^{2}/2Q)$),
spanning 55 msec with 9 oscillations; a numerical simulation of a
black hole binary merger (total mass of 20 solar masses ($M_{\odot}$),
(hereafter referred to as short BH merger)\cite{baker2006bbh}, which
shows strong non-linear frequency modulation from $\sim$300 Hz to
$\sim$900 Hz over 5 msec and 3 oscillations; and a 60 $M_{\odot}$
BH merger (hereafter referred to as long BH merger), drifting in frequency
from $\sim$100 to $\sim$300 Hz over 20 msec. We place these signals
in a time window of 62.5 msec white Gaussian noise, bandpass limited
to 1000 Hz at SNR=8, at a sampling frequency of 16384 Hz.

\section{Detection METHODS\label{sec:Detection-strategies}}

\subsection{Detection statistics and Bayesian blocks }

\begin{figure*}
Sine-Gaussian

\includegraphics[scale=0.35]{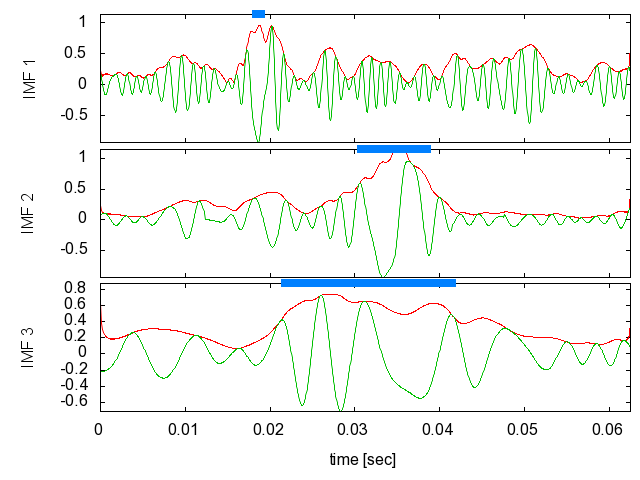}\includegraphics[scale=0.35]{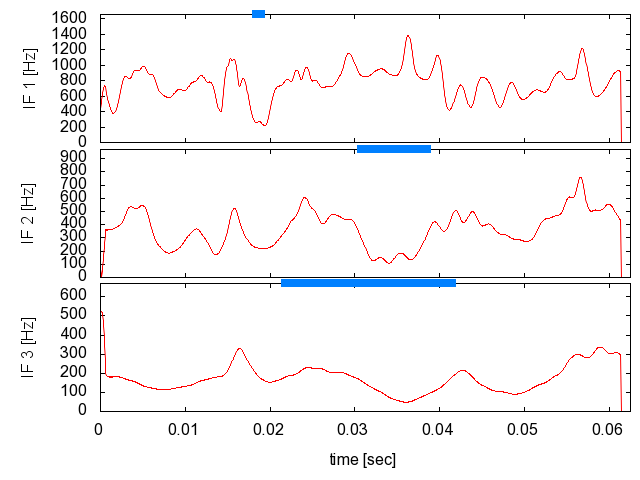}

BH merger short

\includegraphics[scale=0.35]{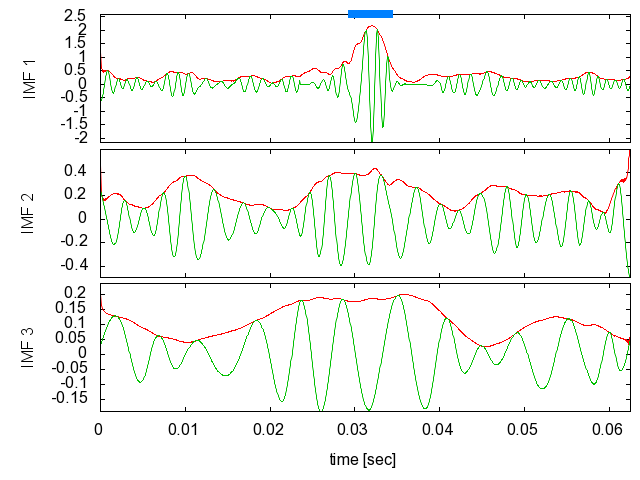}\includegraphics[scale=0.35]{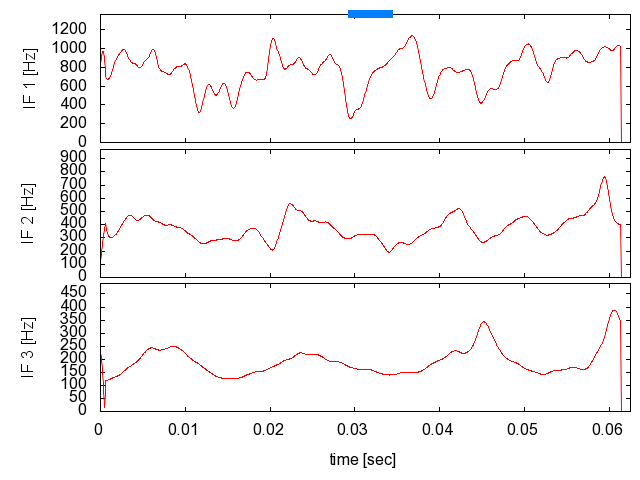}

BH merger long

\includegraphics[scale=0.35]{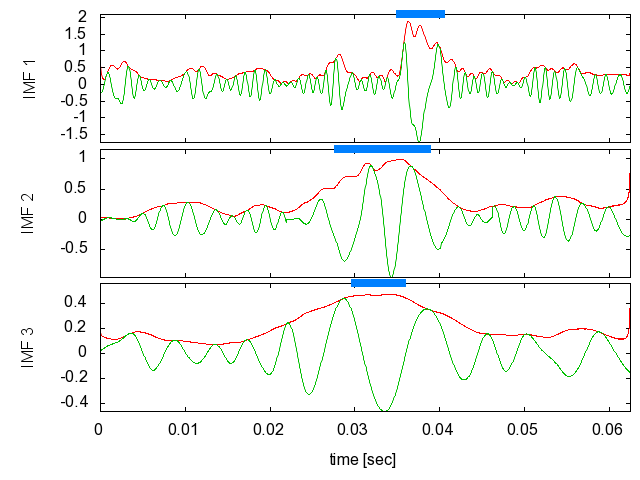}\includegraphics[scale=0.35]{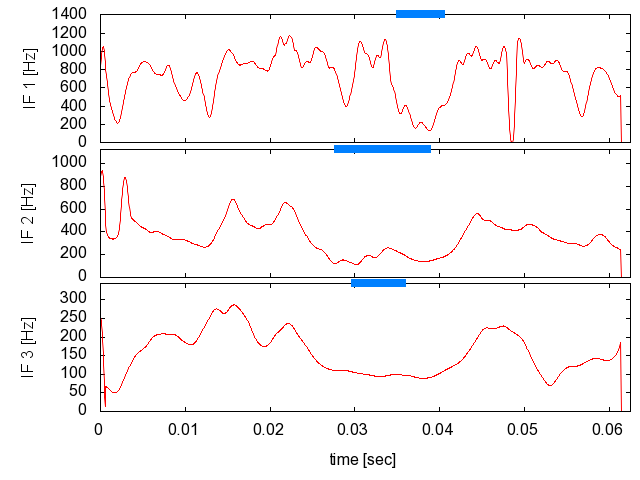}

\caption[IA, triggered blocks]{ The IMF, IA (upper envelope of the IMF) and IF of an EMD decomposition
of our three test signals. The presence of a signal is indicated by
an elevated power level. We observe Bayesian blocking to identify
equi-statistical regions of the IP, with regions in which the mean
and the standard deviation of the IP is significantly different compared
to a noise-only test scenario indicated in blue at the top axis of
each panel. }

\label{f1} 
\end{figure*}

EMD decomposes stationary white noise data in the absence of signals
according to a dyadic filter bank, where the mean frequency of the
IFs in $n$th-IMF is $\propto f_{s}/2n+1$ where $f_{s}$ is the data
sampling rate \cite{flandrin2004emd,huang2005hht,wu2004scw}. The
HHT power spectrum, derived by plotting the instantaneous power IP
( = IA$^{2}$) per frequency interval, shows a uniform distribution
of power over frequencies \cite{wu2004scw}, indicating that the decomposition
preserves the flat frequency content of white noise. This allows us
to formulate an HHT detection strategy for a signal in noise, by searching
for a temporal region of excess power which is statistically distinguishable
from the case of noise with no signal. We show in Fig. \ref{f1} the
EMD decomposition of the three test signals, where the IA is seen
as an upper envelope to the IMFs. Regions of excess power can clearly
be seen.

The excess power in a time series can be quantified in a number of
different ways, each of which provides a statistic which may be compared
to a threshold to indicate a detection. We tested several quantities
for the exact formulation of the statistic: 1) the sum over all the
IAs in every IMF (sumPow), 2) the maximum IA as seen across all IMFs
(maxPow), 3) identifying temporal regions in IA which are significantly
above noise levels in their mean value, and deriving the mean over
these regions (stat1), 4) identifying temporal regions in IA which
are significantly above noise levels in their mean value, and deriving
the sum over these regions (stat2), 5) combining stat1 and stat2 by
adding the normalized statistic values for each trial (stat1+2). We
note that the latter three statistics are adaptive, and require the
use of a technique known as Bayesian blocks \cite{scargle1998sat,mcnabb2004obe}. 

Bayesian blocks determine temporal regions of excess power based on
the Bayesian analysis of the relative probability of two different
hypotheses. The first hypothesis ($M_{1}$) is that a data segment
$X_{n}$ is drawn from a distribution characterized by a single mean
$(\mu)$ and variance $(\sigma^{2})$ and the second, that the segment
$X_{n}$ consists of two continuous and adjacent sub-segments each
drawn from a distribution characterized by a different $\mu$ and/or
$\sigma^{2}$, as separated by a discrete {}``change point'' in
the individual statistics. The probability that a given data set is
drawn from a normal distribution with unknown $\mu$ and $\sigma^{2}$
is equal to:

$P(X_{n}|M_{1})=\int d\sigma\int d\mu(2\pi\sigma^{2})^{-\frac{N}{2}}P(\mu,\sigma)\Pi_{k=0}^{N-1}e^{-\frac{(x_{k}-\mu)}{2\sigma^{2}}}$,

where $P(\mu,\sigma)$ is the a priori probability that the mean and
the variance takes on a specific value. Data blocks are defined by
finding the temporal start and end point of a deviation in $\mu$
or $\sigma$ (in which several change points in one data stretch are
found by iteratively applying the algorithm to each data stretch to
the left and right of a just found change point). Temporal regions
of excess power are defined by finding blocks with significantly elevated
$\mu$ or $\sigma^{2}$. We refer to \cite{mcnabb2004obe} for details.
The identification of the Bayesian blocks that indicate regions of
$4\sigma$-excess power in our test signals is shown in Fig.~\ref{f1}
as the blue bars at the top of the panels.

\subsection{Efficiency Curves}

\begin{figure}
\includegraphics[scale=0.35]{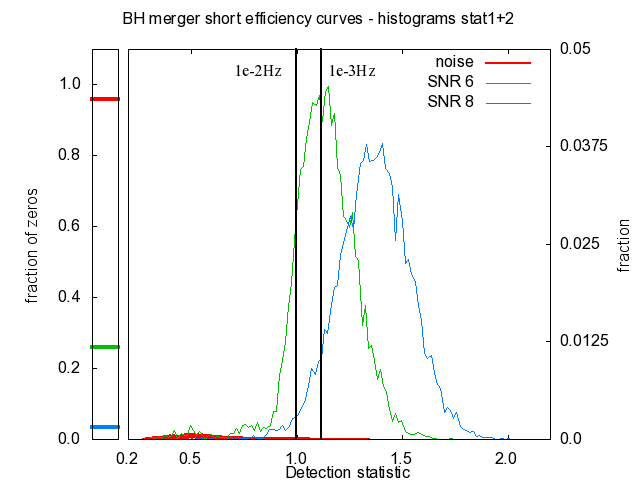}

\includegraphics[scale=0.35]{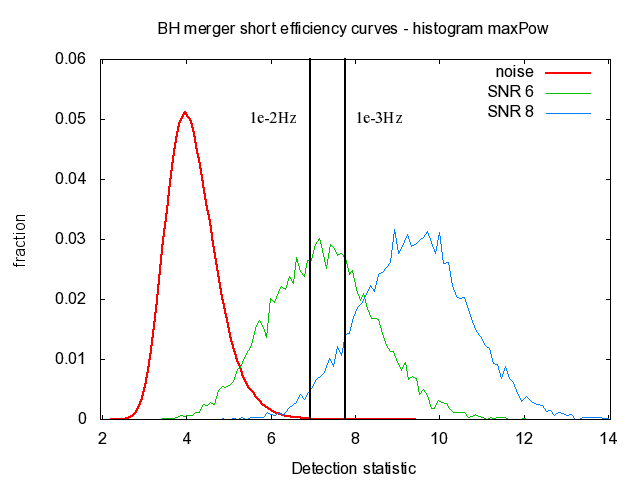}

\caption{We show the histograms of the detection statistic value stat1+2 and
maxPow over $10^{6}$ trials (noise only) and $10^{4}$ trials (signal
injections) for the short BH merger. As stat1+2 implements a threshhold
in IA we find a significant fraction of the detection statistic equal
to zero, as indicated at the left of the plot. The threshhold for
a FAR $10^{-2}$ Hz and $10^{-3}$ Hz is indicated as black vertical
line. Detection efficiency is defined as fraction of triggers above
threshold for given FAR. (For details see text).}

\label{Flo:ROC-1} 
\end{figure}

\begin{figure}
\includegraphics[scale=0.35]{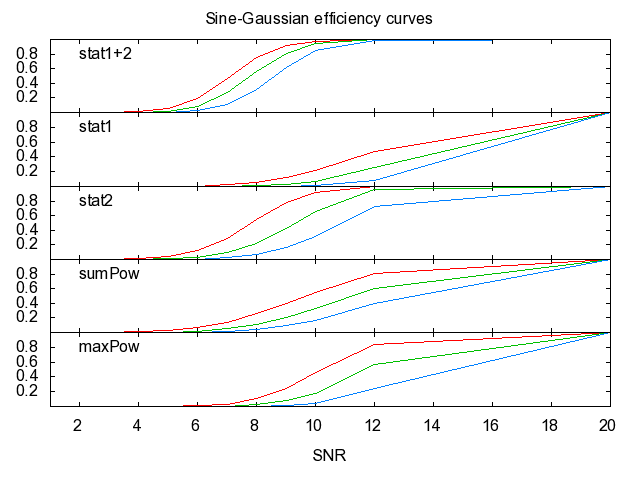}

\includegraphics[scale=0.35]{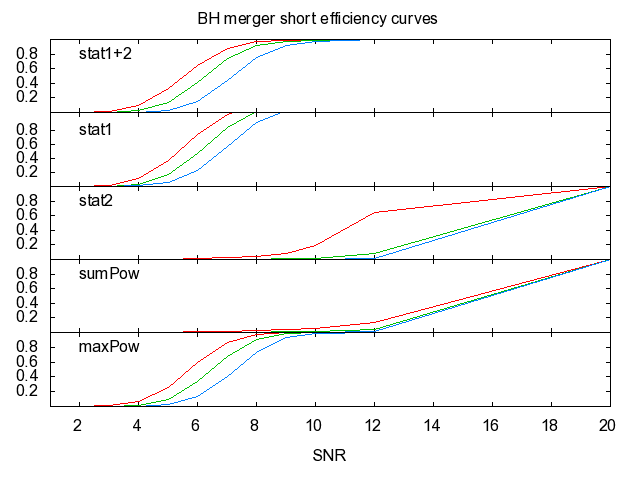}

\includegraphics[scale=0.35]{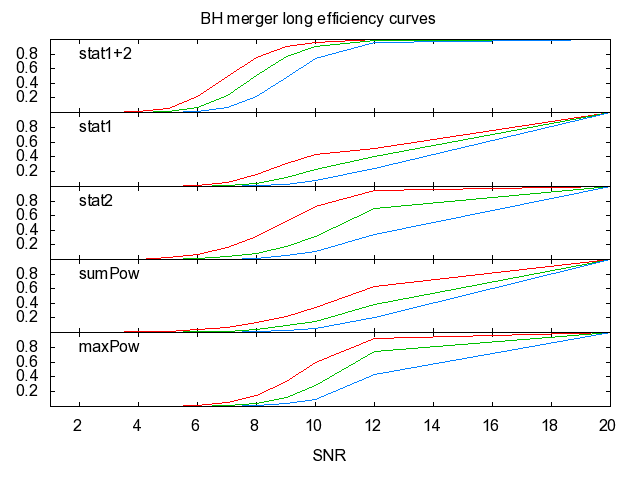}

\caption{The efficiency curves for our three test signals. We show FAR of $10^{-2}$
Hz (red), $10^{-3}$ Hz (green) and $10^{-4}$ Hz (blue), and display
detection efficiency vs SNR. Based on these plots, we choose stat1+2
as our detection statistic.}

\label{Flo:ROC} 
\end{figure}

\begin{figure}
{\resizebox{\hsize}{!}{\includegraphics{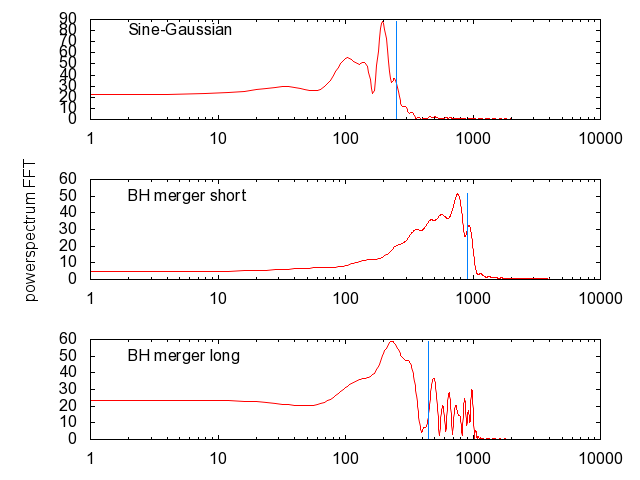}}}

\caption[IA, max freq]{Estimation of the maximum frequency in our three test signals (SNR
8). The generation of a Fourier power spectrum over the boundaries
of the triggered blocks within the lowest IMF allows an estimate of
the signal maximum frequency. The estimate is inexact (high) due to
time-frequency uncertainty of the Fourier analysis. By restricting
the analyzed region to the detected blocks within IMFs we eliminate
a significant fraction of the noise background, allowing a cleaner
determination of the maximum frequency.}

\label{f2-1} 
\end{figure}

To judge the usefulness of the various statistics to identify signals
in noise, we use a Monte Carlo approach to generate detection efficiency
curves as follows\cite{nla.cat-vn1199308,metz1978bpr}. After choosing
the detection statistic, a set of trials are performed on noise without
a signal to track the rate of noise detections, or false alarm rate
(FAR), which may be caused by extrinsic uncertainties or statistical
fluctuations of the noise power at any given instant. A threshold
is set to yield the FAR at a low value, typically $10^{-2}$ Hz to
$10^{-3}$ Hz (for our trial time window of 62.5 msec). Next a signal
of given SNR is added to a white noise time series. The HHT decomposition
is run and the extracted value of the detection statistic is compared
to the threshold to determine whether a detection was made. This is
repeated for a large number of trials to examine the efficiency of
signal detection at a given SNR and FAR. We show in Fig. \ref{Flo:ROC-1}
examples of the histograms for stat1 and maxPow detection statistics
with $10^{6}$ noise-only trials and $10^{4}$ signal injection trials
per given SNR for the short BH merger. We additionally show the noise
only distribution and display the specific FAR of $10^{-2}$ Hz and
$10^{-3}$ Hz, corresponding to 625 and 63 false alarms per million
trials respectively. We note that the adaptive detection statistics
stat1, stat2 and stat1+2 use a $4\sigma$ IA threshold in each decomposed
IMF to identify a block, which is also required to consist of at least
4 data points; these conditions limit noise fluctuations from exceeding
the detection threshold. Thus for stat1+2 $96\%$ percent of noise
detection statistic values are zero, as indicated in Fig. \ref{Flo:ROC-1}.
In contrast, the added amplitude of signals results in an increased
detection efficiency so that only $25\%$ of the detection statistic
values are zero at SNR 6 and $3\%$ are zero at SNR 8. The detection
statistic maxPow and sumPow do not use a threshold in IA, thus their
histograms of the statistics of signal and noise do not show zeros
and are Gaussian in shape. 

We display in Fig. \ref{Flo:ROC} the efficiency curves for all five
detection statistics as applied to the three test signals, where we
plot detection efficiency for the three test signals vs. SNR, with
FAR of $10^{-2}$ Hz, $10^{-3}$ Hz and $10^{-4}$ Hz. We draw a number
of conclusions from these plots. In general we find that shorter signals
at a fixed SNR show a higher detection efficiency, as they show larger
$I\! A/\sigma_{n}$, as discussed earlier. We also find that the adaptive
statistics stat1, stat2 and stat1+2 perform better than maxPow and
sumPow in most cases, due to the efficiency of the Bayesian blocks
in identifying signals. We find that stat1 tends to trigger on excess
power in short regions, while stat2 is more efficient for excess power
over longer intervals, assuming the same amount of total added power
is placed in either short or long regions respectively; thus the combined
statistic stat1+2 appears to be efficient for the range of our test
signals. We identify outliers from extrinsic HHT uncertainties yielding
a $4\sigma$ level IA increase over a given region to primarily contribute
to false alarms in stat1, stat2, and stat1+2. Finally, we note that
all three signals are best detected at the 50$\%$ efficiency level
with stat1+2 for all false alarm rates. We therefore choose stat1+2
as detection statistic for the remainder of the paper.

\subsection{Additional information from Bayesian blocks}

Additional information about the signal can be obtained by analyzing
the detected blocks: the start and end times of the event can be derived
by looking at the outermost edges of all the triggered blocks over
the IMFs, and the SNR of the detected event may be estimated by summing
over the amplitude of the triggered blocks and comparing to the standard
deviation of the noise. Finally, the maximum frequency of the signal,
a very important quantity in this analysis, may be obtained by examining
the Fourier power spectrum of the shortest data stretch that includes
all the triggered blocks of the lowest IMF. We show below that the
signal can be greatly clarified in the characterization stage by applying
a low-pass filter based on this maximum frequency, which removes noise
in the time series that is outside the bandwidth of the signal evolution.
We illustrate the derivation of the maximum frequencies in Fig.~\ref{f2-1},
where we have generated the power spectra of the detected blocks for
each test signal. We find that an effective way to separate the upper
edge of the power spectrum from the noise background is to identify
the first inflection point after the spectrum peak and round up to
the next highest multiple of 50 Hz in frequency. We note that restricting
the analyzed region to the detected blocks eliminates a significant
fraction of the noise background (roughly the ratio of length of the
signal to the length of the time window), allowing a cleaner determination
of the maximum frequency.

\section{Characterization methods}

With the identification of a region of excess power that indicates
a signal, we consider methods to accurately characterize the signal
including its frequency evolution in time. With these methods we seek
to reduce extrinsic and intrinsic uncertainties, and also to estimate
the remaining uncertainty. We further look into the special case of
very low SNR ($<$5) signal characterizations.

\subsection{Reduction of extrinsic and intrinsic uncertainty}

\begin{figure}
\resizebox{\hsize}{!}{\includegraphics{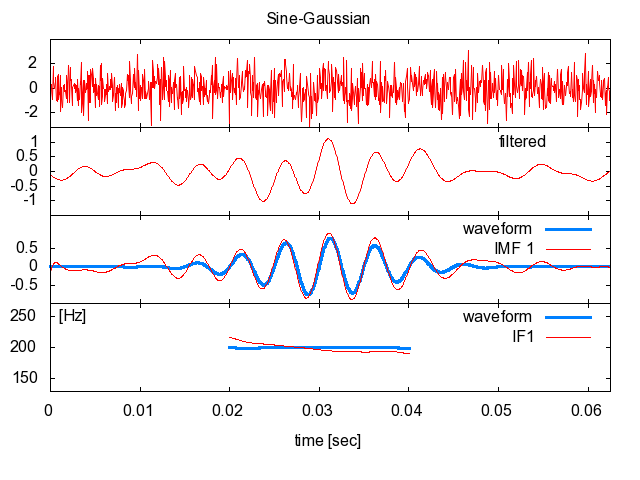}}

\resizebox{\hsize}{!}{\includegraphics{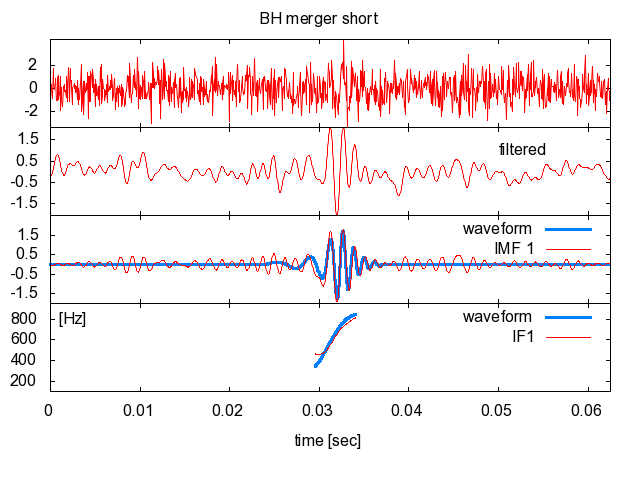}}

\resizebox{\hsize}{!}{\includegraphics{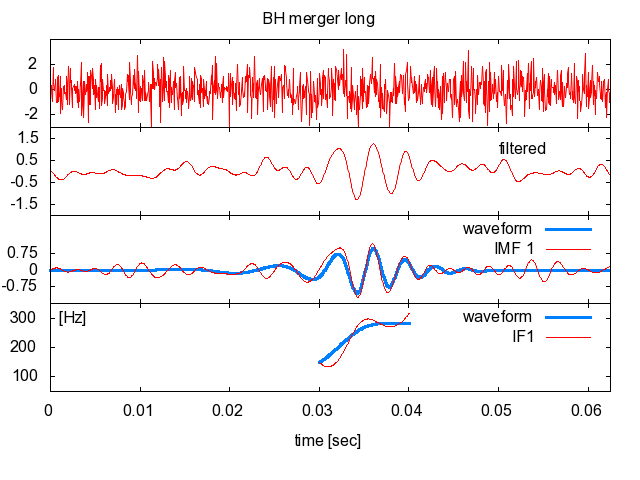}}

\caption[agfilt]{The use of a zero-phase filter specified in the detection stage is
used to remove noise from the data above the maximum frequency of
the event. Here we show the three test signals before (first panel
of each plot) and after the filtering (second panel of each plot).
The third panel shows the result of applying EEMD to the filtered
waveform, which now yields the signal in the first IMF, as opposed
to the multiple IMFs of \ref{f1}. Comparison of this result with
the signal waveform (in blue) shows the effectiveness of the filtering
method. Finally, in the forth panel we show the IFs derived from IMF1,
which are considerably more accurate than those of Fig. \ref{f1}. }

\label{f3} 
\end{figure}

\begin{figure}
\resizebox{\hsize}{!}{\includegraphics{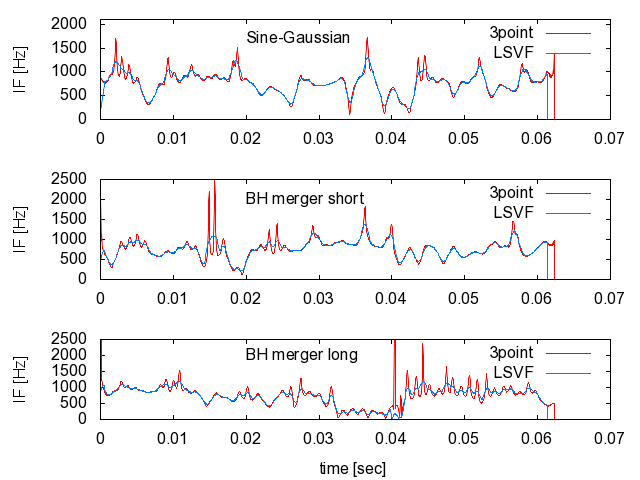}}

\caption[freqfluct]{The IF of the first IMF of the test signals as derived either by
a 3-point differentiation rule or the least squares velocity filter
(LSVF), here using a 20-point averaging. The 3-point estimate experiences
fluctuations that are non-physical, reaching values larger than 1000
Hz in noisy regions although the data bandwidth is 1000 Hz. The LSVF
reduces the largest fluctuations and singularities. }

\label{f2.5-1} 
\end{figure}

The most effective method to suppress extrinsic uncertainties in the
decomposition of the time series into its IMFs was the implementation
of an aggressive zero-phase FIR low-pass filter technique\cite{smith:sae},
removing noise with frequency content above the maximum frequency
of the signal. We illustrate this procedure in Fig.~\ref{f3} using
the test signals. The application of a filter with low-pass frequency
close to the maximum frequency of the waveform removes the need for
decomposition of the data into multiple IMFs before the signal is
encountered, and reduces the associated extrinsic uncertainties caused
by numerous envelope generations, and their possible over-/undershoot.
Using the filtering, the very first EMD decomposition directly targets
the frequency scales of the signal. Furthermore, the filter suppresses
fluctuating noise which can cause different parts of the signal to
appear in different IMFs. The filter needs to satisfy several requirements:
a) it must be zero-phase so that the slope of the signal is not altered
by phase dispersion; b) it should have a very sharp transition from
the pass band (below the filter frequency) to the stop band (above
the filter frequency) to allow it to remove frequency content close
to the signal; c) it should suppress ringing in the pass band and
stop band; d) the filter frequency should be as close as possible
to the maximum frequency of the signal to be most effective in filtering
noise (we estimate that the filter cut off frequency should not be
closer than 50 Hz to the intrinsic maximum frequency of the signal
to avoid distorting the signal waveform.) The filter cut-off frequency
can be found adaptively, as mentioned above, by generating a Fourier
spectrum of the signal region of the lowest IMF identified by the
Bayesian blocks. This estimate yields 250 Hz, 900 Hz and 450 Hz as
filter frequency for the Sine-Gaussian, the short BH merger and the
long BH merger respectively, as shown in Fig.~\ref{f2-1}.

After application of the zero-phase filtering, envelope over/undershoot
in the remaining EMD decomposition can be reduced with ensemble EMD
(EEMD,\cite{WuEEMD2008}) methods. In this technique ensembles of
data are produced by injecting small, random distributions of Gaussian
white noise into the original data stream. EMD is then performed on
each ensemble member, with the result that each member suffers a perturbation
of the position of local maxima and minima and thus a slightly different
envelope fitting. The error propagation related to any member of the
ensemble tends to be attenuated once the ensembles are averaged, yielding
a better measure of the signal compared to one EMD run alone. We average
over 20 ensemble members, and inject white Gaussian noise at 10 percent
of the original time series noise standard deviation.

To demonstrate the effectiveness of the zero-phase filtering, we compare
in Fig.~\ref{f3} the true IF evolution of the test signals with
the decomposed IFs version. The panels of the plots, from top to bottom,
show: 1) the time series of the signal in noise with SNR=8; 2) the
time series obtained from the application of the zero-phase filter
with filter frequency found as described above; 3) the first IMF obtained
from the application of EEMD to the filtered time series; 4) the IF
obtained from the first IMF. In panels 3 and 4, the blue trace shows
the first IMF and IF of the actual signal without noise. We find the
decomposed IF to closely approximate the true IF in all three test
cases: where the filtered signal amplitude is largest the agreement
is $\Delta f/f\leq0.1$, while the error becomes larger at the signal
boundaries where the amplitude becomes comparable to the noise. These
errors represent the combination of remaining intrinsic and extrinsic
uncertainties, and are quantified in the next section. The improvement
of the IF evolution as found in the final characterization stage relative
to the detection stage (Fig. \ref{f1}) is apparent: the Sine-Gaussian
now appears well resolved in its (flat) IF evolution, and the long
BH merger shows significantly less oscillatory behaviour in the IF
after filtering. The short BH merger IF did not change significantly
as the signal showed sufficient $I\! A/\sigma$ contrast to provide
an accurate derivation of the IF in the detection stage, and the data
cut off frequency of 1000 Hz is close to the estimated signal maximum
frequency of 900 Hz.

Remaining uncertainties in the IF determination can take the form
of singularities and/or oscillations outside the physical IF range
which are caused by noise-induced spurious oscillations in the phase.
These oscillations may be controlled by implementing a least squares
velocity filter (LSVF), which provides a 2nd order polynomial fit
to a specified number of points of the IFs, thereby averaging over
sharp IF fluctuations. Frequencies in IMFn that are much greater or
much less than the range implied in the dyadic structure $\propto f\! s/2n+1$
will be attenuated with this technique. In our examples we use a filter
of order 20 to derive the frequency from the phase, which smooths
the shape of 20 consecutive data points to derive one IF value. This
is illustrated in Fig.~\ref{f2.5-1}.

\subsection{Estimation of uncertainties}

\begin{figure}
\resizebox{\hsize}{!}{\includegraphics{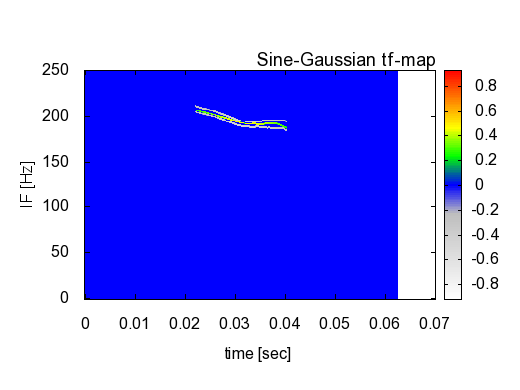}}

\resizebox{\hsize}{!}{\includegraphics{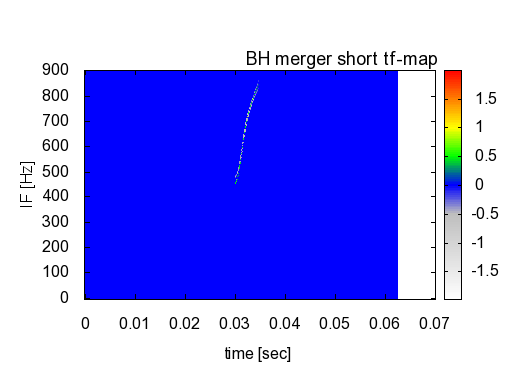}}

\resizebox{\hsize}{!}{\includegraphics{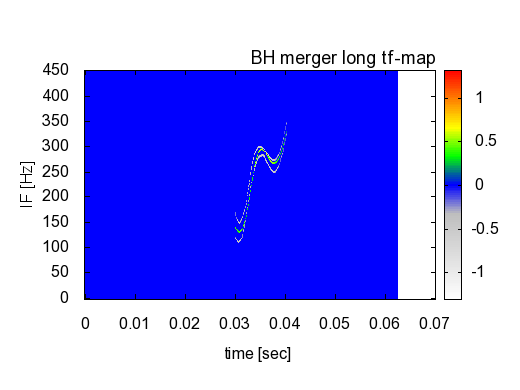}}

\caption[tf-KD-tf-err]{Estimation of the time-frequency-uncertainty for the test signals
with the Sine-Gaussian in the upper panel and the short and long black
hole mergers in the lower two panels. The uncertainty is indicated
by an envelope (white lines) about the time-frequency trace of the
signal (colored trace) }

\label{f5} 
\end{figure}

An analytical estimate of the uncertainty of the IMF, IA or IF of
a decomposed signal at low SNR is difficult to establish, due to the
empirical nature of the HHT decomposition. But it is possible to define
the relative uncertainty of the individual decomposition by looking
at its variance with respect to a perturbation of the time series.
As described above, we use the EEMD averaging process to obtain an
accurate measure of the IFs by averaging and smoothing out extrinsic
uncertainties. We can also use the EEMD as a tool to determine the
uncertainty of the decomposition. Since EEMD injects noise at all
frequencies into the data stream, it will alter both the envelope
fitting and the signal waveform itself. Thus to quantify the total
uncertainty we apply EEMD with an injection of an additional amount
of noise equal to the noise level of the data.

In detail, we create an ensemble consisting of 40 members by summing
the time series with different realizations of white Gaussian noise
with the same standard deviation of noise found in the data (the noise
standard deviation is estimated outside the block boundary which marks
the signal region.) We then apply a zero-phase low-pass filter at
the maximum frequency of the signal to each member of the ensemble,
perform an EMD decomposition and finally derive the IF distribution.
This new decomposition is subject to large intrinsic uncertainties
as the additional injected noise significantly changes the slope of
the waveform, and large extrinsic uncertainties as the EMD envelope
fitting is distorted by the altered local maxima which propagates
through the iterative process. The 40 members of the ensemble then
carry a different IF value at each time; the spread is used to derive
a 2$\sigma$ uncertainty envelope. The IF trace derived from the original
data lies within the uncertainty envelope, with the width of the envelope
displaying the confidence of the IF derivation. In Fig.~\ref{f5}
we show two dimensional time-frequency (tf) maps that plot IF and
IA of the three test signals vs. time, with power as color. We find
the time-frequency traces to fit inside the uncertainty envelopes,
with the uncertainties becoming relatively large where the amplitude
to noise ratio becomes small.

\subsection{Kernel density estimates and very low SNR signal characterization}

\begin{figure*}
\includegraphics[scale=0.4]{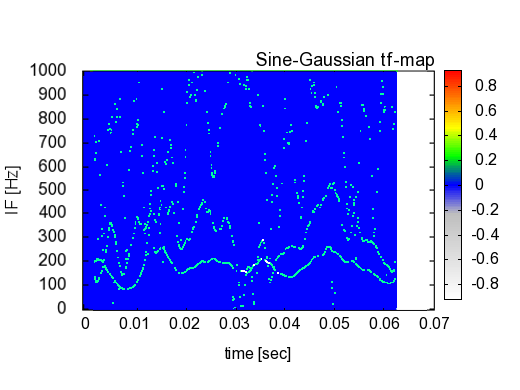}\includegraphics[scale=0.4]{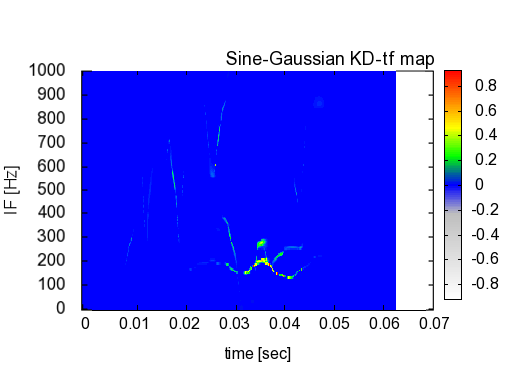}

\includegraphics[scale=0.4]{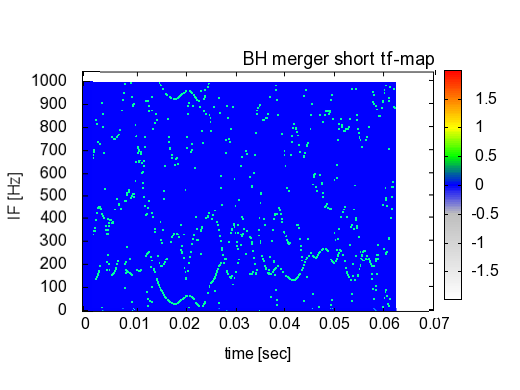}\includegraphics[scale=0.4]{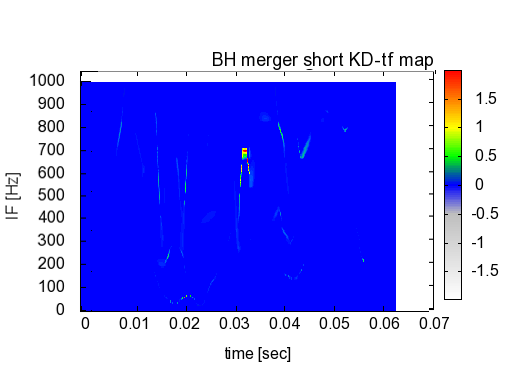}

\caption{The tf-map and the KD-tf map of the Sine-Gaussian at SNR=4 (top panels)
and the short BH merger at SNR=3.}

\label{Flo:KD-tf} 
\end{figure*}

We now consider the case of data analysis on signals at SNR $<$5,
in the case of our test signals $I\! A/\sigma$ ratios of less than
0.6. In order to detect these signals the threshold of the detection
statistic stat1+2 as described in Sec. \ref{sec:Detection-strategies}
has to be lowered (see Fig. \ref{Flo:ROC}). This results in the unavoidable
generation of errors: noise bursts can trigger blocks, resulting in
poorer overall detection sensitivity as noise enters the detection
statistic; the number of false alarms increases; and within the event
blocks we find poorer timing estimates and upper frequency estimates
for the signal. These errors lead to poor tf map evaluations as the
noise may be comparable or stronger in power than the signal, and
also noise is not effectively removed from the tf map by considering
only triggered blocks.

A method to regain signal contrast in SNR$<$5 tf maps is the following.
We seek to take advantage of the fact that signal remnants show coherent
structure and follow the underlying trend of the true signal in the
tf-plane by clustering around its idealized true positions (in absence
of noise) with elevated power levels. Noise in comparison is incoherent
and scatters over the tf plane with random power levels. Thus signal
regions can be identified by locating clustered tf traces with elevated
power levels.

We find weighted kernel density estimates on the tf-plane (KD-tf)
best suited to highlight signal regions \cite{silverman1986des}.
The KD-tf starts with the assumption that a signal structure is evident
in the data, and that it would, in absence of noise, place a coherent
and continuous trace in the tf-plane. We posit this idealistic trace
as a probability density distribution, effectively showing no density
outside the signal trace and a sharp, peaked probability density within
the signal trace where a coherent structure can be found.

We implement an adaptive weighted kernel density estimate to recover
this signal probability density. The kernel, $K(t,f)$, is a bi-variate
Normal distribution, with dimensions of time and frequency \begin{equation}
K(t_{i}-t_{j},f_{i}-f_{j})=\frac{1}{2\pi\sigma_{t_{j}}\sigma_{f_{j}}\sqrt{1-\eta_{j}^{2}}}\exp\left[-\frac{z}{2(1-\eta_{j}^{2})}\right]\label{eq:wkde1}\end{equation}
 with \begin{equation}
z=\frac{\left(t_{i}-t_{j}\right)^{2}}{\sigma_{t_{j}}^{2}}+\frac{\left(f_{i}-f_{j}\right)^{2}}{\sigma_{f_{j}}^{2}}-\frac{2\eta_{j}\left(t_{i}-t_{j}\right)\left(f_{i}-f_{j}\right)}{\sigma_{t_{j}}\sigma_{f_{j}}}\label{eq:wkde2}\end{equation}
 and \begin{equation}
\eta_{j}=\frac{\sigma_{t_{j},f_{j}}}{\sigma_{t_{j}}\sigma_{f_{j}}}\label{eq:wkde3}\end{equation}

In the density estimate every point $[t_{j},f_{j}]$ in the tf-plane
carries its own kernel, whose standard deviation in both dimensions
$\sigma_{t_{j}},\sigma_{f_{j}}$, and correlation $\sigma_{t_{j},f_{j}}$,
is estimated adaptively on the basis of the tf-plane population. Intuitively,
a point within a sparse population (e.g. incoherent noise scattering)
should carry a wide and flat kernel, while points within a coherent
clustering of signal remnants should carry a peaked and sharp kernel.
This contrast allows the signal to be emphasized relative to the noise.
The contrast can be seen most clearly at frequencies near the data
sampling rate, and is found by applying EMD to each IF and retaining
only the first IMF, which removes all but the highest frequency content
in each IF. Scanning each high frequency IF distribution for changes
in the standard deviation, and using the Bayesian blocking technique
to find the kernel specifics at each point, we find signal remnants
and noise sources blocked separately and associated with different
mean, correlation and standard deviation. Each point within a block
uses the statistics of the block to establish its individual kernel.
The final estimate is built according to \begin{equation}
KD\!-\! tf_{i}=\sum_{j}I\! P_{j}^{2}K(t_{i}-t_{j},f_{i}-f_{j})\label{eq:wkde4}\end{equation}
 where we weigh the power of the signal trace in quadratic terms to
balance contributions from clustering (as contained in the Kernel)
with power.

We demonstrate the advantages of the KD-tf for low SNR in Fig. \ref{Flo:KD-tf}.
We show two examples, the Sine-Gaussian at SNR=4 and the short BH
merger at SNR=3. In both cases the estimation of the upper frequency
of the signal was not possible because of the low SNR, and thus filtering
was not used, leading to relatively large extrinsic uncertainties.
We see that the KD-tf maps help to better define the signal remnants
and improve the signal to noise contrast ratio. We find the Sine-Gaussian
to oscillate around its center frequency, and the short BH merger
to only reveal the very peaked part of its waveform, the merger/ring-down
transition at $\sim$870 Hz. We estimate that KD-tf techniques are
applicable only if the maximum $I\! A/\sigma$ ratio is larger than
0.3.

\section{Discussion}

We presented in this paper methods to enable the HHT to efficiently
detect and accurately characterize signals in the low SNR ($<$20)
regime. Since the overall power contained in the data is roughly preserved
by the HHT decomposition, we were able to construct strategies to
search for excess power to find a signal in comparison to a noise
only decomposition. The concept of Bayesian blocking was introduced
to adaptively locate regions of excess power, and therefore to localize
and analyze the signal while also gaining sensitivity by limiting
noise contributions to the detection statistic. We derived efficiency
curves which showed that our black hole merger and Sine-Gaussian test
signals could be detected with better than 50\% efficiency at SNR
$<$8 with FAR = $10^{-3}$ Hz. We then considered measures to accurately
derive the IF at low SNR. We proposed zero-phase FIR low-pass filters
and LSVF filters to reduce the amount of extrinsic IF uncertainties,
and KD-tf mappings to limit intrinsic IF uncertainties. A measure
of the uncertainty or reliability of the decomposition was obtained
by testing the dependence of the extracted signal waveform on increased
extrinsic and intrinsic IF uncertainties.

One direction for future research is the estimation of the upper frequency
of the event. We found FFT methods to overestimate an upper limit
to the signal frequency because of inherent FFT time-frequency uncertainty.
Also, if the signal maximum frequency shows a strong gradient the
FFT estimate becomes even more spread out, resulting in ineffective
subsequent FIR filtering. A possible approach to obtain a more accurate
measure is to assess the maximum frequency with the IF and the IA,
but this is more noise sensitive than the FFT, especially if the signal
shows its largest frequency at low amplitude. A combination of both
approaches might be optimal.

Finally, we have investigated other adaptive decomposition methods
to compare their extrinsic error generation with those of EMD. (ITD\cite{frei2007its},
sawtooth transform\cite{lu2007fim}, fastEMD\cite{luFEMD2008}). We
find that the EMD has a significant advantage due to the way it handles
waveform inflections, which can either be caused by multiple oscillations
needing further decomposition, or by a frequency modulation that is
a true physical property of the underlying waveform. EMD is able to
effectively differentiate between these, while the ITD and the sawtooth
transform methods do not, and the fastEMD incorrectly decomposes all
inflections into multiple oscillations in distinct modes. We note
that only with a proper decomposition can our proposed additional
filters help significantly in the process of detecting and characterizing
the signal.
\begin{acknowledgments}
We acknowledge helpful comments on this manuscript by Peter Shawhan.
This work has been supported in part by NSF grant PHY-0738032. 
\end{acknowledgments}
\bibliographystyle{apsrev} 
\bibliography{hhtatlowsnr}

\end{document}